\documentclass[a4paper, 12pt]{article}
\usepackage[dvips]{graphicx}

\def\ltap{\ \raise.3ex\hbox{$<$\kern-.75em\lower1ex\hbox{$\sim$}}\ }
\def\gtap{\ \raise.3ex\hbox{$>$\kern-.75em\lower1ex\hbox{$\sim$}}\ }
\def\gl{\ \raise.4ex\hbox{$>$\kern-.75em\lower1ex\hbox{$<$}}\ }

\begin{document}

\title{Quark condensates in the chiral bag with the NJL interaction}
\author{S.~Yasui$^a$, A. Hosaka$^b$ and M.~Oka$^a$ \\
\normalsize $^a$Department of Physics, Tokyo Institute of Technology,\\
\normalsize H-27, Meguro, Tokyo 152-8551, Japan \\
\normalsize $^b$Research Center for Nuclear Physics (RCNP), Osaka University,\\
\normalsize Mihogaoka 10-1, Ibaraki, Osaka 567-0047, Japan}
%\date{\today}
\maketitle

\begin{abstract}
We discuss the quark condensate of the vacuum inside the baryon.
We analyze the 1+1 dimensional chiral bag in analogy with the realistic 1+3 dimensional one.
The Nambu--Jona-Lasinio (NJL) type interaction is used to investigate the quark condensate in the chiral bag.
Considering the strong meson-quark coupling, we solve the mean field solution to the scalar and pseudoscalar channels.
Extracting the finite values of the chiral Casimir energy and the quark condensate by proper regularization,
the resulting self-consistent equation allows 
 a generation of a finite dynamical quark mass inside the bag.
%An approximate Cheshire Cat picture holds for massive quarks.
\end{abstract}

\section{Introduction}

The QCD vacuum is a non-perturbative system, and makes it difficult to study the physics of the strong interaction.
Since the early stage of the study of QCD, the MIT bag model has been one of popular models of hadrons \cite{MITbag}.
There, the inside of the bag is assumed to be a perturbative space, and the quark dynamics is treated perturbatively with much success \cite{DeGrand_etal_75}.
Furthermore by requiring chiral symmetry, the bag model was shown to have a pion cloud \cite{Chodos, Inoue}.
The chirally symmetric coupling between pions and quarks at the bag surface gives a conservation of the axial vector current.
However, the strong pion-quark coupling causes an instability of the bag itself \cite{Brown_Rho1979, Brown1979, Vento80}.
In order to avoid this problem, a Skyrmion was introduced outside of the bag, whose model setup is called as the chiral bag model.
Other extensions, such as the cloudy bag model \cite{Thomas}, the chiral bag model with vector mesons \cite{Hosaka90}, are also useful models.

In the QCD vacuum, chiral symmetry is spontaneously broken and the constituent quark mass is induced dynamically \cite{Nambu}.
The non-perturbative structure of the vacuum in the bag is a long standing problem \cite{Milton, Johnson, Kunihiro1, Kunihiro2, Zahed85, Kochelev_85, Dorokhov_etal_92}.
Contrary to the original expectation, it is interesting to see that the vacuum structure inside the bag may become nontrivial, where quark and gluon condensates may take finite values due to the boundary conditions.
In the chiral bag, the pion cloud is also a source for non-perturbative nature inside the bag, since the pions and quarks interact with each other strongly \cite{Zahed85}.
For instance, the quark scalar condensate has a finite value $\langle \bar{\psi}\psi \rangle \simeq - (0.1 \mbox{GeV})^{3}$ in the MIT bag model \cite{Milton} and also in the chiral bag model \cite{Zahed85}.
It is comparable to the observed value $- (0.25 \mbox{GeV})^{3}$ as given in the Nambu--Jona-Lasinio (NJL) model \cite{Nambu} and the other approaches.
Therefore, it would be a natural question whether the non-perturbative vacuum  is induced by the boundary conditions or some dynamical interactions such as the NJL one.   

In this paper, to understand the non-perturbative dynamics in the bag, the NJL interaction is introduced as a quark-quark interaction inside the chiral bag \cite{Kunihiro1, Kunihiro2}.
This approach is justified when the color confinement is caused by long range gluon dynamics at distance $\simeq 1$ fm, while the chiral symmetry breaking occurs at shorter distance scale $\simeq 0.2- 0.3$ fm, as suggested by instanton dynamics \cite{Kochelev_85, Dorokhov_etal_92, Birse, Manohar, Kunihiro85}.
Indeed such a separation of the length scales has been considered in the context of the NJL model in the hadron dynamics \cite{Kunihiro85}.
In recent approaches based on AdS/CFT correspondence, it is suggested that there is a window of the gauge coupling where chiral symmetry breaking occurs in a deconfinement phase \cite{Bak_Yee}.

Further analysis for the hybrid model of the NJL model and the chiral bag model, which may be called as the NJL chiral bag model, was given in \cite{Kiriyama_Hosaka, Yasui1, Yasui2, Yasui3, Yasui4}.
It is well known that the strong pion-quark coupling in the chiral bag causes the chiral vacuum polarization inside the bag \cite{Chodos, Inoue, Brown_Rho1979, Brown1979, Vento80, Brown_Rho88, Hosaka_Toki92, Brown84, Mulders, Hosaka_Toki86, Hosaka_Toki96, Hosaka}.
Familiar examples are the conservation of baryon number and the chiral Casimir energy \cite{Zahed85, Goldstone, Zahed_Meissner84}.
It was shown that the NJL chiral bag with finite quark mass also holds the properties of the vacuum polarization \cite{Farhi, Yasui4}.

In the previous studies of the NJL bag model \cite{Kiriyama_Hosaka, Yasui1, Yasui2, Yasui3, Yasui4}, only the scalar condensate was considered.
However, this channel causes a divergence of the chiral Casimir energy in the limit of zero bag radius.
In the present work, we will show that a self-consistent solution can be found with finite quark condensates for the baryon number $B=1$ system by considering not only the scalar channel but also the pseudoscalar channel.
The latter channel is required by the hedgehog ansatz in the pion sector.
In this way, we can study the quark condensates inside hadrons (quark bag) by both quark boundary conditions (chiral Casimir effects) and the self-consistency condition from the NJL interaction.
%In this way, we can study the effects of both quark boundary conditions (chiral Casimir effects) and the NJL interaction which cause finite quark condensates inside hadrons (quark bag).
We evaluate the quark condensates as mode sums in the chiral bag model without introducing a cutoff as in the ordinary NJL model.
We consider this as a counter representation of the long-range
dynamics with a cutoff, which is valid inside the chiral bag.

In this paper, in order to avoid numerical complications of the 1+3 dimensional model, instead, 
we consider a 1+1 dimensional system with the Gross-Neveu model \cite{Gross_Neveu_74} in analogy with the realistic case.
In this way, we expect to learn essential non-trivial dynamics of the NJL interaction inside the bag.
In our model, quarks are confined in a finite segment of the line in a U(1) $\times$ U(1) chirally symmetric way.
This simplification has an advantage that an analytical calculation can be performed.
Such a model setup does not modify the qualitative feature of the chiral vacuum polarization in the realistic 1+3 dimensional  bag.
Indeed, in the conventional chiral bag model with massless quarks, the 1+1 dimensional chiral bag was shown to provide a clear understanding of the quark vacuum polarization properties \cite{Zahed84, Zahed_Klabucar}.

This paper is organized as follows.
In Section 2, the NJL chiral bag model is introduced in the 1+1 dimensional system.
Using the hedgehog ansatz and the mean field approximation for the NJL interaction, a self-consistent equation is derived.
Chiral vacuum polarizations such as the baryon number conservation and the chiral Casimir energy are investigated carefully.
The self-consistent equation is solved and the total energy of the NJL chiral bag is investigated.
In Section 3, our results are discussed.
The final section is devoted to the conclusion.

\section{The chiral bag with the NJL interaction in one dimension}

\subsection{Lagrangian}

The purpose of this section is to formulate the 1+1 dimensional chiral bag with the NJL \cite{Nambu} (or the Gross-Neveu \cite{Gross_Neveu_74}) interaction.
Here, we discuss effects of finite quark mass, which are induced by the mean field of the NJL interaction, to the vacuum polarization in the chiral bag.
We consider the lagrangian
\begin{eqnarray}
{\cal L} &=&
  \left[ \bar{\psi} 
 \frac{i}{2} \stackrel{\leftrightarrow}{\partial} \hspace{-0.35cm}/ \, \psi
  - G \left\{ (\bar{\psi}\psi)^{2}+(\bar{\psi}i \gamma_{5} \psi)^{2} \right\} \right]  \theta(R-|x|)
              - \frac{1}{2} \bar{\psi} U^{\gamma_{5}} \psi \delta(|x|-R)
 \label{eq : lagrangian} \\ 
    &&+ \left[ -\frac{1}{2} (U^{\dag}\partial_{\mu}U)(U^{\dag}\partial^{\mu}U) - \frac{\lambda^{2}}{2} (2 - U - U^{\dag}) \right] \theta(|x|-R),
     \nonumber
\end{eqnarray}
where the quark field $\psi$ has single flavor with U(1) $\times$ U(1) symmetry inside the one dimensional bag $|x|<R$.
In 1+1 dimension, the quark field is expressed by a two component Dirac spinor with the gamma matrices given as
\begin{eqnarray}
\gamma_{0} = 
\left(
\begin{array}{cc}
  0 & 1 \\
  1 & 0    
\end{array}
\right),
\gamma_{1} = 
\left(
\begin{array}{cc}
  0 & -1 \\
  1 & 0    
\end{array}
\right),
\gamma_{5} = 
\left(
\begin{array}{cc}
  1 & 0 \\
  0 & -1    
\end{array}
\right),
\end{eqnarray}
in the chiral representation.
The quarks inside the bag interact with each other through the NJL interaction in the second term in the first bracket with a coupling constant $G$.
The second term with the $\delta$ function represents an interaction between quarks and pions at the bag surface at $|x|= R$.
The last term is the meson lagrangian outside the bag $|x|>R$.
We use the sine-Gordon field with U(1) $\times$ U(1) symmetry \cite{Zahed84}
\begin{eqnarray}
  U = e^{i \phi},
\end{eqnarray}
which mimics the topological property of the pion field in the 1+3 dimensional system in the Skyrme model \cite{Skyrme, Adkins, Zahed_Brown86}.
The delta function with $U^{\gamma_{5}} = e^{i \gamma_{5} \phi}$ gives a chirally symmetric interaction between quark and pion at the bag boundary.
The last term with $\lambda$ is a cosine potential to give a dynamically stable soliton solution, in which $\lambda$ plays a role of the ``pion mass".
The pion decay constant is a dimensionless quantity in the 1+1 dimensional system, and can be eliminated by rescaling the chiral field and the pion mass.
The energy of the sine-Gordon field is $8 \lambda$ when we take the zero bag radius limit.

\begin{figure}[ptb]
\begin{center}
\includegraphics[width=8cm, angle=0, clip]{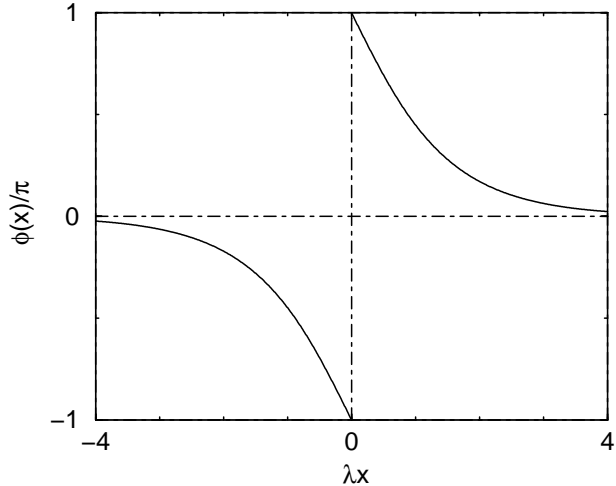}
\end{center}
\vspace*{0.0cm} \caption{\small \baselineskip=0.5cm The solution of the sine-Gordon equation in the zero bag radius limit. See Eq.~(\ref{eq : sineGordon_solution}).}
 \label{fig : sineGordon}
\end{figure}

Considering the strong coupling between quarks and mesons at the bag surface, we introduce the ``hedgehog" mean fields.
In the meson field for $|x| \ge R$, we consider
\begin{eqnarray}
 \phi(x) = \epsilon(x)F(x),
\end{eqnarray}
with a chiral angle $F$ which is a positive function of the position $x$.
The sign function $\epsilon(x) = x/|x|$ represents the ``hedgehog" structure in the 1+1 dimensional system \cite{Zahed84}.
The equation of motion of the meson,
\begin{eqnarray}
 \partial_{t}^{2} \phi - \partial_{x}^{2} \phi - \lambda^{2} \sin(\phi-\pi) = 0,
\end{eqnarray}
has a static solution in the limit of the zero bag radius,
\begin{eqnarray}
  F(x) = \epsilon(x) \left( 2\pi - 4 \arctan e^{\lambda |x|} \right).
  \label{eq : sineGordon_solution}
\end{eqnarray}
Note that $F(x)$ approaches zero in the limit $|x| \rightarrow \infty$ as shown in Fig~\ref{fig : sineGordon}.
This behavior mimics the Skyrmion solution in the 1+3 dimension.
Our solution coincides with the conventional solution in the sine-Gordon equation except for a phase factor.
For instance, if we replace $F(x) \rightarrow F(x) + 2 \pi$ for $x<0$, we find a continuous solution at $x=0$ satisfying $F(-\infty)=2\pi$ and $F(+\infty)=0$.

%%%%%%%%%%%%%%%%%%%%%%%%%%%%%%%%%%
%%%%%%%%%%%%%%%%%%%%%%%%%%%%%%%%%%

For the quark sector inside the bag, we introduce quark condensates not only in the scalar channel but also in the pseudoscalar channel.
In the mean field approximation, we take the following ansatz
\begin{eqnarray}
 -2G \langle \bar{\psi} \psi \rangle &=& m \cos F,
 \label{eq : mean_field1} \\
 -2G \langle \bar{\psi} i \gamma_{5} \psi \rangle &=& m\, \epsilon(x) \sin F,
\label{eq : mean_field2}
\end{eqnarray}
for a self-consistent equation with a dynamical quark mass $m$.
The mean field approximation is taken in a basis set of the quark wavefunctions in the bag.
Here we assume that the quark mass $m$ and chiral angle $F$ are chosen to be independent of the position $x$ inside the bag.
We mention that $F$ is continuous to $F(x)$ at $|x|=R$.
As a result, 
we obtain the lagrangian
\begin{eqnarray}
{\cal L} &=& 
\bar{\psi} \left[ \frac{i}{2} \stackrel{\leftrightarrow}{\partial} \hspace{-0.35cm}/ 
             - m e^{i  \gamma_{5}F \epsilon(x)} \right] \psi \theta(R-|x|)
              - \frac{1}{2} \bar{\psi} e^{i  \gamma_{5} F \epsilon(x)} \psi \delta(|x|-R)
   \label{eq : lagrangian2} \\
    &&+ \left[ \frac{1}{2} (\partial_{x} F)^{2} - \lambda^{2} \left( 1+\cos(F-\pi) \right) \right] \theta(|x|-R).
    \nonumber
\end{eqnarray}
The quark mass term with a constant chiral angle $F$ keeps a symmetry of the lagrangian
 under the transformation $F \rightarrow \pi-F$ as we see below.
We call the term, $e^{i  \gamma_{5} F \epsilon(x)}$, as the chiral phase.
The chiral phase was equal to one in our previous analysis, since only the scalar channel was chosen in the mean field approximation \cite{Yasui4}.
There, the chiral Casimir energy diverges in the small bag radius limit.
In the following discussion, we will show that the chiral phase plays an important role, not only in the vacuum polarization, but also in the quark condensates.

The surface term in Eq.~(\ref{eq : lagrangian2}) induces a bag boundary condition at $x=R$
\begin{eqnarray}
 i\gamma_{1}\psi = - e^{iF\gamma_{5}} \psi.
\end{eqnarray}
Then, the eigenvalue equation is obtained as 
\begin{eqnarray}
 E(1+\kappa \sin F) \sin kR - \kappa \cos F (k \cos kR + m \sin kR) = 0,
 \label{eq : eigen1}
\end{eqnarray}
for the states with $|E| \ge m$ with quark energy $E^{2} = k^{2} + m^{2}$ and momentum $k$.
For  $|E| < m$, replacing $k \rightarrow i k$ with $E^{2}=-k^2+m^2$, we obtain
\begin{eqnarray}
 E (1+\kappa \sin F) \sinh kR - \kappa \cos F (k \cosh kR + m \sinh kR) = 0.
 \label{eq : eigen2}
\end{eqnarray}
The parity $\kappa=\pm1$ is defined by the parity transformation
\begin{eqnarray}
 \psi(x) \rightarrow \gamma_{0} \psi(-x) = \kappa \psi(x).
\end{eqnarray}

\begin{figure}[ptb]
\begin{center}
\includegraphics[width=10cm, angle=0, clip]{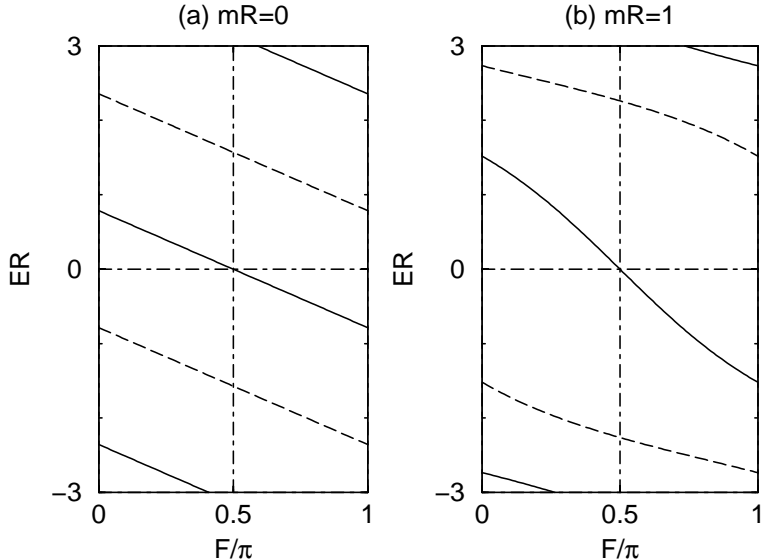}
\end{center}
\vspace*{0.0cm} \caption{\small \baselineskip=0.5cm
Quark eigenenergies as functions of the chiral angle $F$ for (a) $mR=0$ and (b) $mR=1$.
The solid and dashed lines indicate the parity $\kappa=+$ and $-$, respectively.
 }
 \label{fig : ER_F_mR}
\end{figure}

In Fig.~\ref{fig : ER_F_mR}, we plot the quark eigenenergy $E$ as functions of the chiral angle $F$ for massless quark,  $mR=0$, and massive quark, $mR=1$.
It is seen that the quark energies are odd functions of $F-\pi/2$.
Namely the quark spectrum is anti-symmetric under the transformation of $F \rightarrow \pi - F$.
It is also interesting that the energy levels are periodic with the periodicity $\pi$ along with $F$.
In particular, the lowest level crosses $E=0$ at $F=\pi/2$ regardless the mass value $m$.
This is due to the chiral phase in the mass term in Eq.~(\ref{eq : lagrangian2}); without the chiral phase this property is not maintained as shown in \cite{Yasui3, Yasui4}.
In Fig.~\ref{fig : ER_F_mR}, in the small energy region near $E=0$, we see that the spectrum for massive quark is modified as compared with that for the massless quark.
However, the asymptotic behaviors in the high energy region are qualitatively similar to the massless case.

%%%%%%%%%%%%%%%%%%%%%%%%%%%%%%%%%%
%%%%%%%%%%%%%%%%%%%%%%%%%%%%%%%%%%

\subsection{Chiral vacuum polarizations}

In the NJL chiral bag model, various quark matrix elements have contributions from the vacuum due to the modification of the Dirac spectrum by the strong pion field.
In this subsection, we discuss the vacuum polarization including the effect of the finite quark mass.
In principle, the quark mass $m$ is related with the chiral angle $F$ and the bag radius $R$ through the self-consistent equations, (\ref{eq : mean_field1}) and (\ref{eq : mean_field2}).
However, in order to understand the finite quark mass effect on the vacuum polarization, 
we take the quark mass as a constant value irrespective to the chiral angle and the bag radius in this subsection.

Let us first investigate the baryon number.
In the previous studies, the total baryon number was shown to be always conserved, where the contribution of the bag vacuum played an essential role \cite{Goldstone, Zahed_Meissner84, Farhi, Yasui4}.
Here, once again in the present model, the exact conservation of the baryon number can be shown.

The expectation value of the baryon number carried by quarks is defined by the symmetric sum over positive and negative energy states.
The baryon number is defined by
\begin{eqnarray}
   B_{\rm q}(m, F) = - \frac{1}{2} \lim_{\eta \rightarrow +0} \left[ \sum_{E_{n}>0}  e^{-\eta E_{n}R} - \sum_{E_{m}<0}  e^{-\eta |E_{m}|R} \right],
    \label{eq : baryon_number0}
\end{eqnarray}
where the summation converges thanks to the convergence factor $e^{-\eta |E_{n}|R}$.
Instead of the exponential type, it is much easier to use the Strutinsky method in numerical calculations.
This method has an advantage that a sufficient convergence is obtained by summing these series up to $n \ltap 20$ \cite{Wust, Vepstas_Jackson}, while in the other methods we need more states, typically $n \ltap 100$ \cite{Vepstas_Jackson_Goldhaber, Hosaka_Toki}.
The numerical results agree with the form
\begin{eqnarray}
B_{\rm q}(m, F) = 
\left \{
\begin{array}{c}
 -F/\pi  \hspace{0.8cm} \mbox{for} \hspace{0.5cm} 0 \le F < \pi/2 \\
 1-F/\pi \hspace{0.5cm} \mbox{for} \hspace{0.5cm} \pi/2 \le F \le \pi,
\end{array}
\right.
\label{eq : baryon_number}
\end{eqnarray}
which is valid for any quark mass $m$.
This solution is obtained easily by analytical calculation especially for the massless quark.
Therefore, containing the valence quarks with the baryon number $B_{\rm val}=1$ for $0 \le F < \pi/2$ and $0$ for $\pi/2 \le F \le \pi$, we obtain the baryon number in the quark sector as
\begin{eqnarray}
B_{\rm val} + B_{\rm q} =
\left \{
\begin{array}{c}
 1-F/\pi  \hspace{0.5cm} (B_{\rm val}=1) \hspace{0.5cm} \mbox{for} \hspace{0.5cm} 0 \le F < \pi/2 \\
 1-F/\pi \hspace{0.5cm} (B_{\rm val}=0)  \hspace{0.5cm}\mbox{for} \hspace{0.5cm} \pi/2 \le F \le \pi,
\end{array}
\right.
\end{eqnarray}
which give fractional baryon numbers depending on the chiral angle.

The total baryon number in the quark and pion sectors is conserved irrespective to the chiral angle.
We define the pion topological current
\begin{eqnarray}
 J^{\mu} = \frac{i}{2\pi} \varepsilon^{\mu \nu} U^{\dag} \partial_{\nu} U = -\frac{1}{2\pi} \varepsilon^{\mu \nu} \partial_{\nu} \phi.
\end{eqnarray}
Then, the fractional baryon number in the pion sector is
\begin{eqnarray}
 B_{\phi} = \int_{|x|>R} J^{0} {\rm d}x = -2 \frac{1}{2\pi} \int_{R}^{\infty} \frac{dF}{dx} {\rm d}x = F/\pi.
\end{eqnarray}
Then, the total baryon number is given as sum of them in the quark and pion sectors; $B=B_{\rm val}+B_{\rm q}+B_{\phi}=1$.

%The chiral Casimir energy

\begin{figure}[tbp]
\begin{center}
\includegraphics[width=8cm, angle=0, clip]{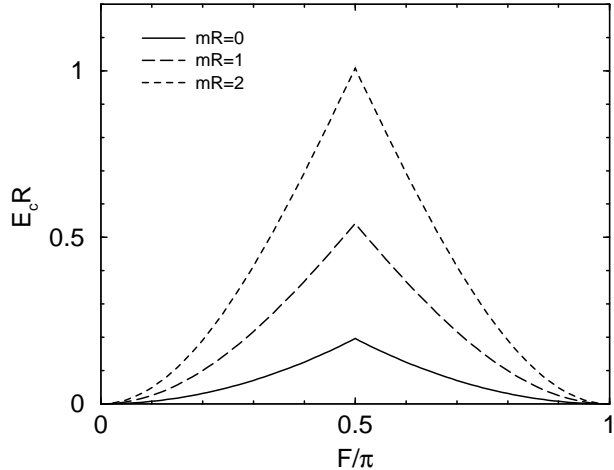}
\end{center}
\vspace*{-1.0cm} \caption{\small \baselineskip=0.5cm The chiral Casimir energy $E_{\rm c}$ as a function of the chiral angle $F$. The solid, long-dashed and short-dashed lines indicate the quark masses $mR=0$, $1$ and $2$, respectively.}
 \label{fig : EcR_F_mR}
\end{figure}

Next we consider the chiral Casimir energy of the bag vacuum.
The chiral Casimir energy is defined as the difference of the energies at $F$ and $F=0$,
\begin{eqnarray}
 E_{\rm c}(m, F) = \tilde{E}_{\rm c}(m,F) - \tilde{E}_{\rm c}(m,0),
\end{eqnarray}
where
\begin{eqnarray}
  \tilde{E}_{\rm c}(m, F) = - N_{\rm c} \frac{1}{2} \lim_{\eta \rightarrow +0} \left[ \sum_{E_{n}>0} E_{n}  e^{-\eta E_{n}R} - \sum_{E_{m}<0} E_{m} e^{-\eta |E_{m}|R} \right],
   \label{eq : chiral_Casimir_energy}
\end{eqnarray}
with $N_{\rm c}=3$.
Especially for massless quarks, an analytical result is obtained as
\begin{eqnarray}
E_{\rm c}(0,F) =
\left \{
\begin{array}{c}
 N_{\rm c}F^{2}/4\pi   \hspace{1.5cm} \mbox{for} \hspace{0.5cm} 0 \le F < \pi/2 \\
 N_{\rm c}(F-\pi)^{2}/4\pi \hspace{0.5cm}\mbox{for} \hspace{0.5cm} \pi/2 \le F \le \pi.
\end{array}
\right.
\end{eqnarray}
In general, the numerical results for the chiral Casimir energies are shown as functions of the chiral angle for several quark masses in Fig.~\ref{fig : EcR_F_mR}.
It is a remarkable point that the chiral Casimir energy vanishes at the chiral angle $F=\pi$, not only for massless quarks, but also for massive quarks.
This is because the chiral phase in the quark mass term in Eq.~(\ref{eq : lagrangian2}) conserves the chiral symmetry in quark sector.
Indeed, the chiral phase guarantees the periodic structure of the quark spectrum and the energy spectrums coincide at $F=0$ and $F=\pi$ except for parity as shown in Fig.~\ref{fig : ER_F_mR}.
This result ensures a continuity from the bag model to the Skyrmion, as we discuss later.
We mention that, without the chiral phase, the chiral Casimir energy takes an infinite value at $F=\pi$ and the continuity is not maintained for massive quarks, as discussed in \cite{Yasui3, Yasui4}.

%%%%%%%%%%%%%%%%%%%%%%%%%%%%%%%%%%
%%%%%%%%%%%%%%%%%%%%%%%%%%%%%%%%%%

\subsection{Self-consistent solutions}

Now, we consider the self-consistent equations (\ref{eq : mean_field1}) and (\ref{eq : mean_field2}).
We regard the quark mass and the chiral angle as averaged values inside the bag, since they are assumed to be independent of the position.
Correspondingly, we also consider that the quark scalar and pseudoscalar densities for each eigenstate $n$ are
 defined as values averaged over the bag volume,
\begin{eqnarray}
 \left[ \bar{\psi}_{n} \psi_{n} \right]_{\rm V} &\equiv& \frac{1}{V} \int_{-R}^{R} \bar{\psi}_{n} \psi_{n} {\rm d}x,
 \\
 \left[ \bar{\psi}_{n} i \gamma_{5} \epsilon(x) \psi_{n} \right]_{\rm V} &\equiv& \frac{1}{V} \int_{-R}^{R} \bar{\psi}_{n} i \gamma_{5} \epsilon(x) \psi_{n} {\rm d}x,
\end{eqnarray}
with bag volume $V=2R$.
Here, $\epsilon(x)$ is multiplied in Eq.~(\ref{eq : mean_field2}) for convenience in the following discussions.

Next, we calculate the vacuum polarization of the scalar and pseudoscalar condensates.
For this purpose we consider the following sum over all the states with positive and negative energies,
\begin{eqnarray}
 \langle \bar{\psi} \psi \rangle_{\rm sea}
  &=& -N_{\rm c} \frac{1}{2} \lim_{\eta \rightarrow +0}
                           \left[ \sum_{E_{n}>0} \left[ \bar{\psi}_{n} \psi_{n} \right]_{\rm V} e^{-\eta E_{n}R} 
                                 - \sum_{E_{n}<0} \left[ \bar{\psi}_{n} \psi_{n} \right]_{\rm V} e^{-\eta |E_{n}|R}  \right],
  \label{eq : sum1} \\
  \langle \bar{\psi} i \gamma_{5} \epsilon(x) \psi \rangle_{\rm sea}
               &=& -N_{\rm c} \frac{1}{2}  \lim_{\eta \rightarrow +0}
                           \left[ \sum_{E_{n}>0} \left[ \bar{\psi}_{n}  i \gamma_{5} \epsilon(x) \psi_{n} \right]_{\rm V} e^{-\eta E_{n}R} 
                                 - \sum_{E_{n}<0} \left[ \bar{\psi}_{n}  i \gamma_{5} \epsilon(x) \psi_{n} \right]_{\rm V} e^{-\eta |E_{n}|R} \right].
 \nonumber \\
\label{eq : sum2}
\end{eqnarray}
The scalar condensate is an odd function of $F-\pi/2$, and the pseudoscalar condensate is an even function.
In other words, the former has an odd symmetry for change $F \rightarrow \pi-F$ and the latter has an even symmetry.
Once the quark condensates are obtained for $0 \le F < \pi/2$, they are applied straightforwardly to $\pi/2 \le F \le \pi$.
Therefore it is sufficient to consider only the range of $0 \le F < \pi/2$ in the followings.

The mode sums (\ref{eq : sum1}) and (\ref{eq : sum2}) contain logarithmic divergences coming from the ultraviolet energy region unlike the baryon number and the chiral Casimir energy.
For massless quarks, we obtain analytically the asymptotic behavior in the limit $\eta \rightarrow +0$,
\begin{eqnarray}
 \langle \bar{\psi} \psi \rangle_{\rm sea}
  &=& \frac{ N_{\rm c}}{2\pi V} \cos F
       \left[ H\left(-\frac{1}{2}-\frac{F}{\pi}\right) + H\left(-\frac{1}{2}+\frac{F}{\pi}\right) + 2 \log \frac{\pi}{2} + 2 \log \eta \right] + {\cal O}[\eta],
 \label{eq : massless1} \\
 \langle \bar{\psi} i \gamma_{5} \epsilon(x) \psi \rangle_{\rm sea}
   &=& \frac{N_{\rm c}}{4\pi V}
       \left[ H\left(-\frac{1}{4}-\frac{F}{2\pi}\right) - H\left(-\frac{1}{4}+\frac{F}{2\pi}\right)
            + H\left(-\frac{3}{4}+\frac{F}{2\pi}\right) - H\left(-\frac{3}{4}-\frac{F}{2\pi}\right)
      \right.
    \nonumber \\
       &&\left.
            +2 \sin F \left\{  H\left(-\frac{1}{2}-\frac{F}{\pi}\right) +  H\left(-\frac{1}{2}+\frac{F}{\pi}\right)
                                   + 2\log \frac{\pi}{2} + 2 \log \eta \right\}
       \right] + {\cal O}[\eta],
    \nonumber \\
       \label{eq : massless2}
\end{eqnarray}
for $0 \le F < \pi/2$.
Here, $H(x)$ is a harmonic number defined as
\begin{eqnarray}
 H(x) = - \frac{\gamma+2\log 2}{\pi} - \frac{2}{\pi} \Psi(x),
\end{eqnarray}
 with the Euler constant $\gamma$ and the digamma function $\Psi(x)$.
It shows that the scalar and pseudoscalar condensates have the logarithmic divergences proportional to $(\cos F \, \log \eta)/\pi$ and $ (\sin F \, \log \eta)/\pi$, respectively.
These terms induce ultraviolet divergence in the quark condensates at high momentum region.
The logarithmic divergences for massive quarks has been also confirmed numerically, provided that the coefficient of the divergent term depends on the quark mass.
Considering the asymptotic form of the divergent terms, we remove the divergences and obtain finite values in the following prescription.
Keeping the discrete symmetry between $F$ and $\pi-F$, we define
%\footnote[1]{Concerning the second term in (\ref{eq : reg2}), the derivative $\left. \partial^{2} \langle \bar{\psi}  i \gamma_{5} \epsilon(x) \psi \rangle_{\rm sea}/\partial F^{2} \right|_{F=\pi}$ is not applied, since the obtained pseudoscalar condensate (including sea and valence quarks) has a cusp at $F=\pi/2$.}
\begin{eqnarray}
\langle \bar{\psi} \psi \rangle_{\rm sea}^{\rm fin}
  &=& \langle \bar{\psi}\psi \rangle_{\rm sea} 
   - \left. \frac{\partial^{2} \langle \bar{\psi} \psi \rangle_{\rm sea}}{\partial F^{2}} \right|_{F=\pi} \cos F,
   \label{eq : reg1} \\ 
\langle \bar{\psi} i \gamma_{5} \epsilon(x) \psi \rangle_{\rm sea}^{\rm fin}
  &=& \langle \bar{\psi} i \gamma_{5} \epsilon(x) \psi \rangle_{\rm sea}
   - \left. \frac{\partial^{2} \langle \bar{\psi} \psi \rangle_{\rm sea}}{\partial F^{2}} \right|_{F=\pi} \sin F.
\label{eq : reg2}
\end{eqnarray}
%This regularization holds the symmetry between $F$ and $\pi-F$ which is possessed
We notice that in (\ref{eq : reg1}) and (\ref{eq : reg2}), the direct subtraction of the divergent part, such as $\langle \bar{\psi}\psi \rangle_{\rm sea} - \langle \bar{\psi}\psi \rangle_{sea, F=0}$, is not appropriate, since it breaks the symmetry between $F$ and $\pi-F$.
The reference point is chosen at $F=\pi$ for both the scalar and pseudoscalar densities.
It is also possible to choose $F=0$ as a reference point, provided that the signs in the second terms in (\ref{eq : reg1}) and  (\ref{eq : reg2}) are changed to plus.
Especially for massless quarks, we obtain the analytical results;
\begin{eqnarray}
  \langle \bar{\psi} \psi \rangle_{\rm sea}^{\rm fin}
   &=& - \frac{N_{\rm c}}{2V} \cos F 
 \left[
  - \frac{2}{\pi} (\gamma + 2 \log 2 )
  - \frac{1}{\pi} \left\{ \Psi \left( \frac{1}{2}-\frac{F}{\pi} \right) + \Psi \left( \frac{1}{2}+\frac{F}{\pi} \right) \right\}
  + \frac{28 \, \zeta (3)}{\pi^{3}}
   \right] ,
   \nonumber \\
  \label{eq : quark_density1} \\
  \langle \bar{\psi} i \gamma_{5} \epsilon(x) \psi \rangle_{\rm sea}^{\rm fin}
 &=& - \frac{N_{\rm c}}{4\pi V} \left[  \Psi \left( \frac{1}{4}-\frac{F}{2\pi} \right)
     	                            -\Psi \left( \frac{3}{4}-\frac{F}{2\pi} \right) 
	                            -\Psi \left( \frac{1}{4}+\frac{F}{2\pi} \right)
	                           +\Psi \left( \frac{3}{4}+\frac{F}{2\pi} \right)   \right] 
\nonumber \\
	&&+ \frac{N_{\rm c}}{2 \pi V}\sin F
	 \left[
	          \Psi \left( \frac{1}{2}-\frac{F}{\pi} \right) + \Psi \left( \frac{1}{2}+\frac{F}{\pi} \right)
              + 2 \gamma + 4 \log 2 -\frac{28\zeta(3)}{\pi^{2}}
	 \right],
\label{eq : quark_density2}
\end{eqnarray}
for $0 \le F < \pi/2$.
Here $\zeta(x)$ is the zeta function.
Using the symmetry properties of the scalar and pseudoscalar condensates, the results can be extended to all values of $F$.
%It is checked numerically that this subtraction scheme is also valid for massive quarks.

%%%%%%%%%%%%%%%%%%%%%%%%%%%%%%%%%
\begin{figure}[tbp]
\begin{minipage}{8cm}
\vspace*{0.0cm}
\centering
\includegraphics[width=7cm]{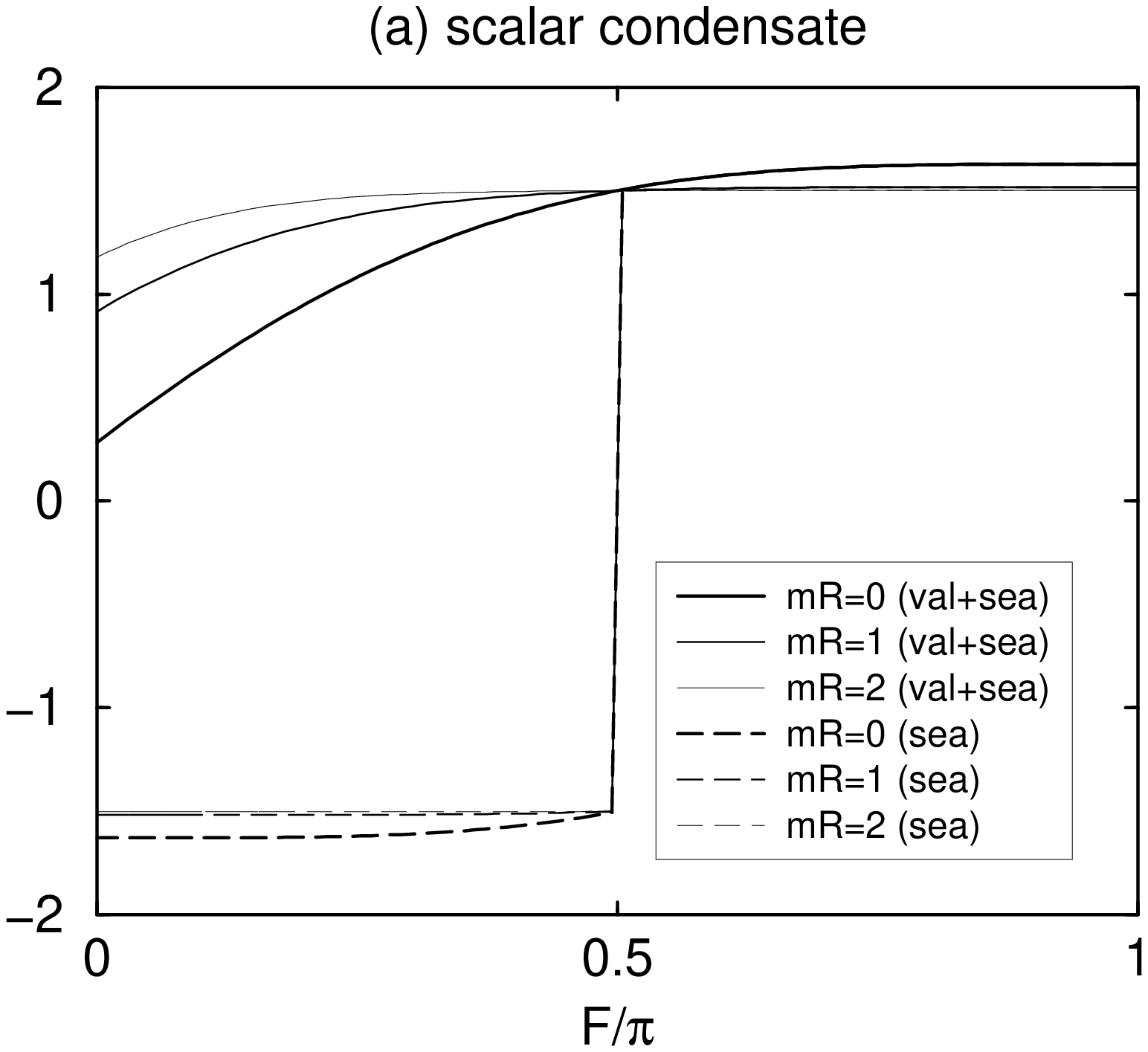}
\vspace{-0.0cm}
\end{minipage}
%%%%%%%%%%%%%%%%%%%%%%%%%%%%%%%%%
\begin{minipage}{8cm}
\centering
\includegraphics[width=7cm]{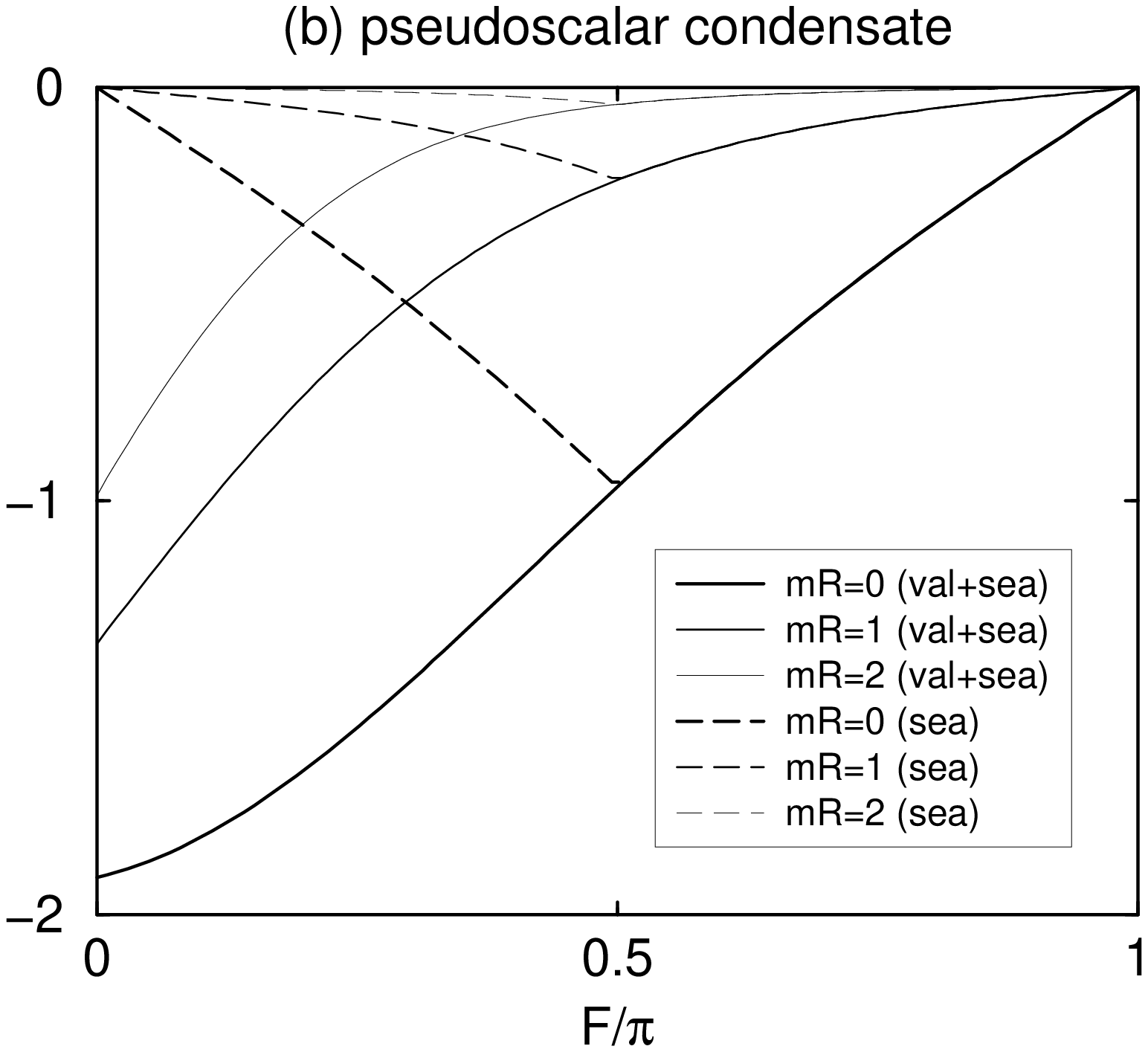}
\vspace{-0cm}
\end{minipage}
\caption{\small \baselineskip=0.5cm The quark condensates in units of $1/V$ as functions of the chiral angle $F$ for several quark masses $mR=0$, $1$ and $2$. (a) the scalar condensate and (b) the pseudoscalar condensate. The solid lines indicate the sum of the valence and vacuum contributions, and the dashed lines indicate only the sea contribution.}
    \label{fig : bqq_mR}
\end{figure}
%%%%%%%%%%%%%%%%%%%%%%%%%%%%%%%%%%%%%%%%%%%%%%%%%%

%Including the valence quarks, 
%we consider a quark condensate $ \langle \bar{\psi} e^{i \gamma_{5} F \epsilon(x)} \psi \rangle$, which is directly related with the self-consistent equation.
%This quantity becomes finite through the following procedure. 
The quark condensate including the valence quarks is given as a sum of valence and sea quark contributions,
\begin{eqnarray}
\langle \bar{\psi} e^{i \gamma_{5} F \epsilon(x)} \psi \rangle^{\rm fin}
= \theta \left( \frac{\pi}{2}-F \right) N_{\rm c} \left[ \bar{\psi}_{\rm val} e^{i \gamma_{5} F \epsilon(x)} \psi_{\rm val} \right]_{\rm V}
+ \langle \bar{\psi} e^{i \gamma_{5} F \epsilon(x)} \psi \rangle_{\rm sea}^{\rm fin}.
\label{eq : quark_density_1}
\end{eqnarray}
The sea quark contribution is obtained after performing the subtraction (\ref{eq : reg1}) and (\ref{eq : reg2}),
\begin{eqnarray}
  \langle \bar{\psi} e^{i \gamma_{5} F \epsilon(x)} \psi \rangle_{\rm sea}^{\rm fin}
  &=&  \langle \bar{\psi} \psi \rangle_{\rm sea}^{\rm fin} \cos F 
        + \langle \bar{\psi} i \gamma_{5} \epsilon(x) \psi \rangle_{\rm sea}^{\rm fin} \sin F.
%\nonumber \\
%  &=&  \langle \bar{\psi} \psi \rangle_{\rm sea} \cos F 
%   + \langle \bar{\psi} i \gamma_{5} \epsilon(x) \psi \rangle_{\rm sea} \sin F
%   -  \left. \frac{\partial^{2} \langle \bar{\psi} \psi \rangle_{\rm sea}}{\partial F^{2}} \right|_{F=\pi}.
 \label{eq : fin}
\end{eqnarray}
The valence quark contribution is
\begin{eqnarray}
\left[ \bar{\psi}_{\rm val} e^{i \gamma_{5} F \epsilon(x)} \psi_{\rm val} \right]_{\rm V}
= \left[  \bar{\psi}_{\rm val} \psi_{\rm val} \right]_{\rm V} \cos F
 + \left[ \bar{\psi}_{\rm val}  i\gamma_{5} \epsilon(x) \psi_{\rm val} \right]_{\rm V} \sin F,
\end{eqnarray}
where $\psi_{\rm val}$ is the wave function of the lowest $0^{+}$ state with positive energy.
In (\ref{eq : quark_density_1}), the valence quark is contained only for $0 \le F<\pi/2$, as indicated by the step function.

In Fig.~\ref{fig : bqq_mR} (a) and (b), the scalar condensate,
\begin{eqnarray}
\theta\left( \frac{\pi}{2}-F \right) N_{\rm c} \left[ \bar{\psi}_{\rm val}\psi_{\rm val} \right]_{\rm V} + \langle \bar{\psi}\psi \rangle_{\rm sea}^{\rm fin},
\end{eqnarray}
and the pseudoscalar condensate,
\begin{eqnarray}
\theta\left( \frac{\pi}{2}-F \right) N_{\rm c} \left[ \bar{\psi}_{\rm val}i\gamma_{5}\epsilon(x)\psi_{\rm val} \right]_{\rm V} + \langle \bar{\psi}i\gamma_{5}\epsilon(x)\psi \rangle_{\rm sea}^{\rm fin},
\end{eqnarray}
in Eq.~(\ref{eq : quark_density_1}) are shown as functions of the chiral angle $F$ for several quark masses $m$, respectively.
For $0 \le F<\pi/2$ the valence quarks are contained, while for $\pi/2 \le F \le \pi$ they are not contained. 
The sum of the valence and sea quarks are indicated by solid lines, and the sea quarks are by dashed lines.
The scalar condensates are positive, while the pseudoscalar condensates are negative.

However, these quark condensates do not supply a continuous transformation to the MIT bag in the large bag radius limit, since the above defined scalar condensate gives a finite value at $F=0$ and $m=0$.
In order to impose of the MIT bag condition, we choose the reference point of the quark condensate at $F=0$,
\begin{eqnarray}
 \langle \bar{\psi} e^{i \gamma_{5} F \epsilon(x)} \psi \rangle^{\rm phys} 
 = \langle \bar{\psi} e^{i \gamma_{5} F \epsilon(x)} \psi \rangle^{\rm fin} 
   - \left. \langle \bar{\psi}  \psi \rangle^{\rm fin} \right|_{F=0, m=0}
   \label{eq : quark_density_2}
\end{eqnarray}
This condition guarantees that for massless quark the quark condensate (\ref{eq : quark_density_2}) becomes zero in the limit $F \rightarrow 0$, and hence quarks become massless in the large MIT bag.
It also results in the condition that the scalar component of the quark condensate vanishes in this limit, 
where the bag boundary conditions play no role for the generation of the  quark condensate.
%Therefore the inside of the large bag is a trivial vacuum.

\begin{figure}[tbp]
\begin{minipage}{8cm}
\vspace*{0.0cm}
\centering
\includegraphics[width=7cm]{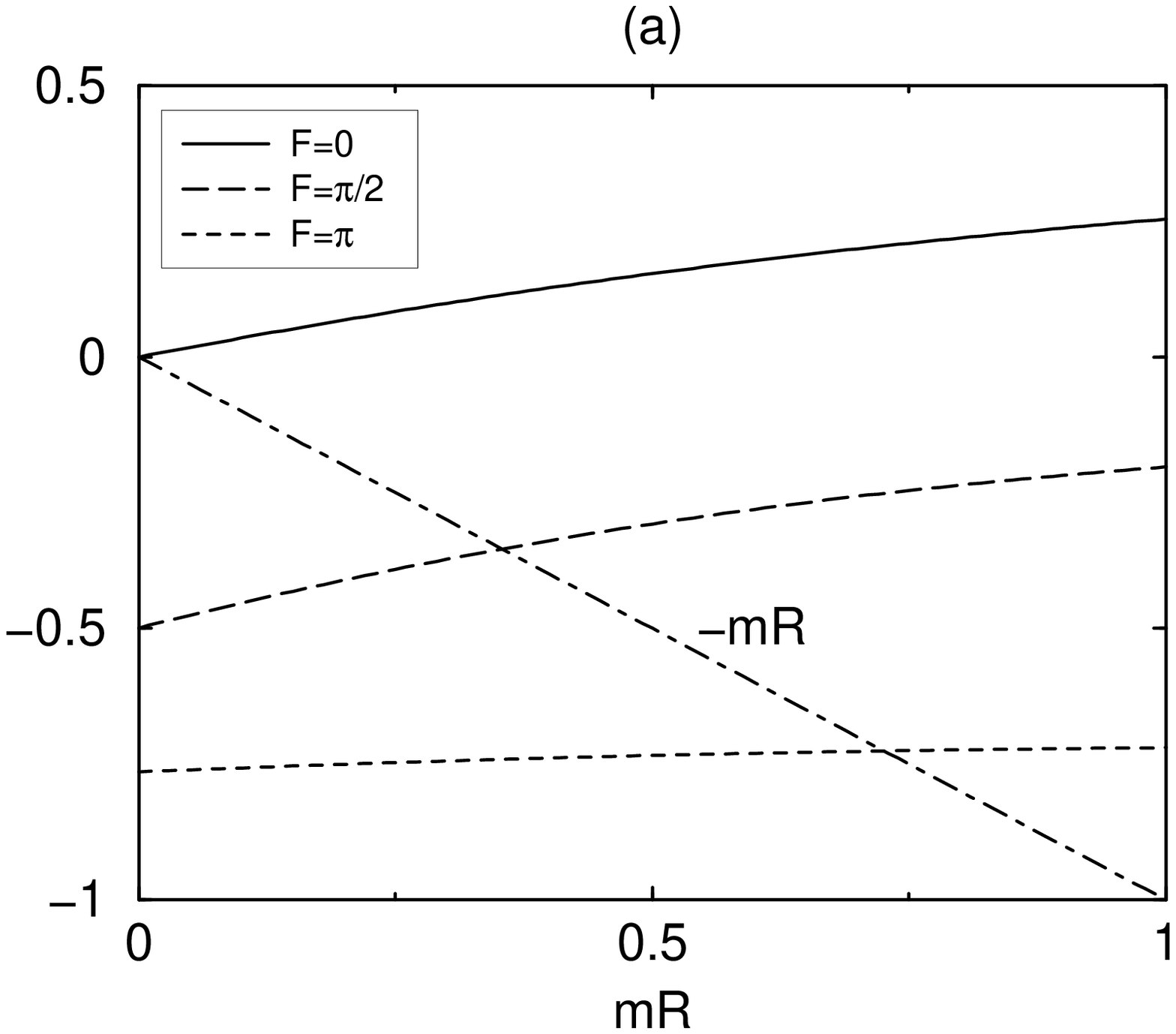}
\vspace{-0.0cm}
\end{minipage}
%%%%%%%%%%%%%%%%%%%%%%%%%%%%%%%%%
\begin{minipage}{8cm}
\centering
\includegraphics[width=7cm]{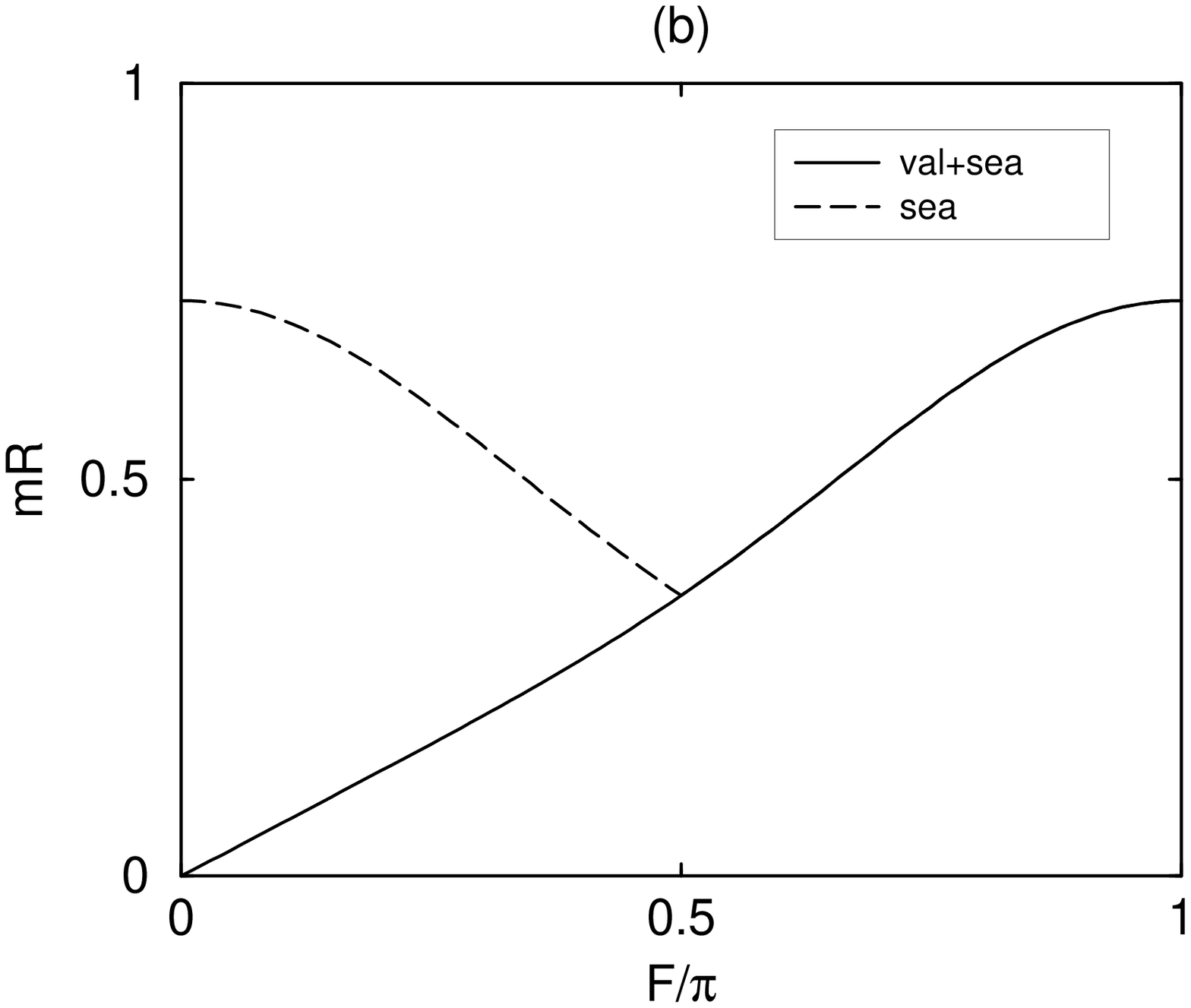}
\vspace{-0cm}
\end{minipage}
\caption{\small \baselineskip=0.5cm 
(a) The quark condensate, $2G R  \langle \bar{\psi} e^{i \gamma_{5} F \epsilon(x)} \psi \rangle^{\rm phys}$, in the the left hand sides in Eq.~(\ref{eq : self2}) as functions of $mR$ for $G=0.2$. See the text for details.
(b) The quark mass $m$ as functions of the chiral angle $F$ for $G=0.2$.
The solid line contains valence and sea quarks.
The dashed line indicates the solution in an empty bag in which the valence quarks are not contained.
}
    \label{fig : mR_F_g}
\end{figure}

Now let us consider the self-consistent equation for dynamical quark mass
\begin{eqnarray}
 -2G R  \langle \bar{\psi} e^{i \gamma_{5} F \epsilon(x)} \psi \rangle^{\rm phys} = mR,
\label{eq : self2}
\end{eqnarray}
which is obtained from the mean field approximation (\ref{eq : mean_field1}) and (\ref{eq : mean_field2}) and the redefinition of the quark condensate indicated in Eq.~(\ref{eq : quark_density_2}).
Here the bag radius $R$ is multiplied in the both sides to make them dimensionless.
By setting $G=0.2$, 
 we compare the left and right hand sides of the self-consistent equation (\ref{eq : self2}) for several chiral angles $F$ and quark masses $m$.
The quark condensate, $2G R  \langle \bar{\psi} e^{i \gamma_{5} F \epsilon(x)} \psi \rangle^{\rm phys}$, in the left hand side is shown as a function of the quark mass for each chiral angle $F=0$, $F=\pi/2$ and $F=\pi$ by the thick solid, long-dashed and short-dashed lines, respectively, in Fig.~\ref{fig : mR_F_g} (a).
The right hand side is indicated by the dot-dashed line in the same figure.
As we see, the chiral angle $F=0$ gives zero quark mass, while the other $F=\pi/2$ and $F=\pi$ give finite quark masses $mR=0.35$ and $mR=0.72$, respectively.

More explicitly, we show the quark mass $m$ as a function of the chiral angle $F$ in Fig.~\ref{fig : mR_F_g} (b).
The solid line indicates the self-consistent solution in the bag with the valence quarks contained, while the dashed line indicates the solution in an empty bag with no valence quark.
The quark mass increases as the chiral angle increases.
It is noted that the quark mass is equal to zero at $F=0$ due to the MIT bag condition (\ref{eq : quark_density_2}).

It is interesting to consider an empty bag, although such a state cannot exist in reality.
There, the quark mass has a maximum value at $F=0$ and $\pi$, and the minimum value at $F=\pi/2$.
This result is interpreted as a change of the quark mass when the pion field moves along the chiral circle of $\sigma^{2}+\pi^{2}={\rm const}$.
It indicates that the quark mass increases as the chiral angle approaches the $\sigma$ axis ($F=0$ and $\pi$), 
while it decreases as the chiral angle approaches the $\pi$ axis ($F=\pi/2$).
On the other hand, once the valence quarks are included for $0 \le F<\pi/2$, the quark condensate is suppressed and the quark mass becomes smaller.
It is consistent with our intuitive understanding about the vacuum.

%%%%%%%%%%%%%%%%%%%%%%%%%%%%%%%%%
%%%%%%%%%%%%%%%%%%%%%%%%%%%%%%%%%%%%%%%%%%%%%%%%%%

\subsection{Energy}

Now, we discuss the total energy of the NJL chiral bag, the sum of the quark and pion energies;
\begin{eqnarray}
 E = E_{\rm q} + E_{\pi} = \theta \left( \frac{\pi}{2}-F \right) E_{\rm val} + E_{\rm c} + E_{\pi},
\end{eqnarray}
with the quark energy $E_{\rm q}$ as a sum of the valence quark energy $E_{\rm val}$ for $0 \le F<\pi/2$, indicated by the step function, and the chiral Casimir energy $E_{\rm c}$, and the pion energy given as
\begin{eqnarray}
 E_{\pi} = 2 \int_{R}^{\infty} \left[ \frac{1}{2} (\partial_{x} F)^{2} + \lambda^{2} (1+\cos F) \right] {\rm d}x.
\end{eqnarray}
For a given bag radius, we solve the equation of motion for pion, and perform the total energy variation with respect to the chiral angle.
This procedure is nothing but the continuity condition for the axial current at the bag surface.
We have two free parameters; the pion mass $\lambda$ and the NJL coupling constant $G$.
We fix $\lambda = 0.125$ GeV to obtain the soliton mass $1$ GeV in the Skyrmion limit.
We use $G=0.2$ and $0$ in the followings.

\begin{figure}[tbp]
\begin{center}
\includegraphics[width=7cm, angle=0, clip]{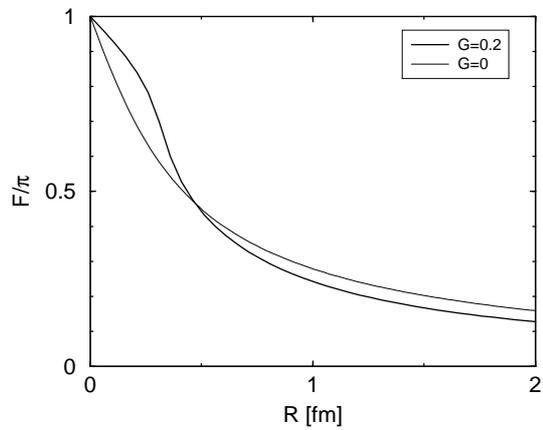}
\end{center}
\vspace*{0.0cm} \caption{\small \baselineskip=0.5cm The chiral angle as functions of the bag radius for $G=0.2$ (thick line) and $G=0$ (thin line).}
 \label{fig : F_R_m}
\end{figure}

The chiral angle is plotted as a function of the bag radius for $G=0.2$ (thick line) and 0 (thin line) in Fig.~\ref{fig : F_R_m}.
The chiral angle approaches $\pi$ in the limit of the small bag radius.
Consequently the Skyrmion and the MIT bag is connected smoothly by changing the bag radius.
%under a continuous transformation.
%This means that the Cheshire Cat picture, which was presented in the conventional chiral bag model, also holds in the NJL chiral bag model.

\begin{figure}[tbp]
\begin{center}
\includegraphics[width=7cm, angle=0, clip]{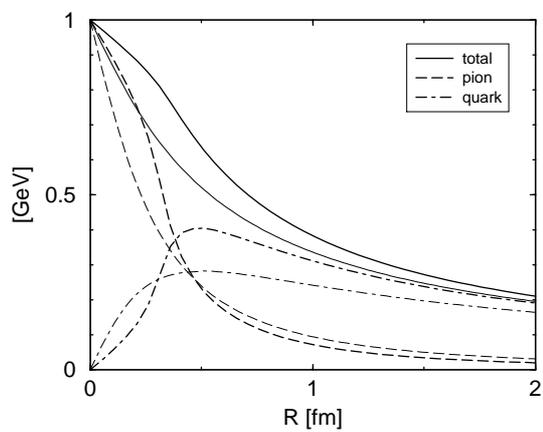}
\end{center}
\vspace*{0.0cm} \caption{\small \baselineskip=0.5cm The total (solid lines), the pion (dashed lines) and quark (dot-dashed lines) energies as functions of the bag radius. $G=0.2$ for thick lines and $G=0$ for thin lines.}
 \label{fig : E_R_m}
\end{figure}

In Fig.~\ref{fig : E_R_m}, we show the total energy $E$ (solid line), pion energy $E_{\pi}$ (dashed line), quark energy $E_{q}$ (dot-dashed line) as functions of the bag radius $R$.
Here $G=0.2$ and 0 are represented by thick lines and thin lines, respectively.
As seen in the conventional chiral bag model ($G=0$), the pion energy is dominant rather than the quark energy for small bag radius, and vice versa for large bag radius.
The former gives the Skyrmion, and the latter gives the MIT bag.
The dominance of each contribution is clearly seen as the NJL coupling constant is switched on; the pion dominates for $R \ltap 0.4$ fm and the quark dominates for $R \gtap 0.4$ fm.
Note that the total energy becomes rather independent of the bag radius $R$ when the volume term $BV$ with a suitable value of the bag constant $B$.
Therefore, the Skyrmion and the MIT bag are smoothly connected by varying the bag radius, consistent with Ceshire Cat picture, which is well known in the conventional chiral bag model.

\begin{figure}[tbp]
\begin{center}
\includegraphics[width=7cm, angle=0, clip]{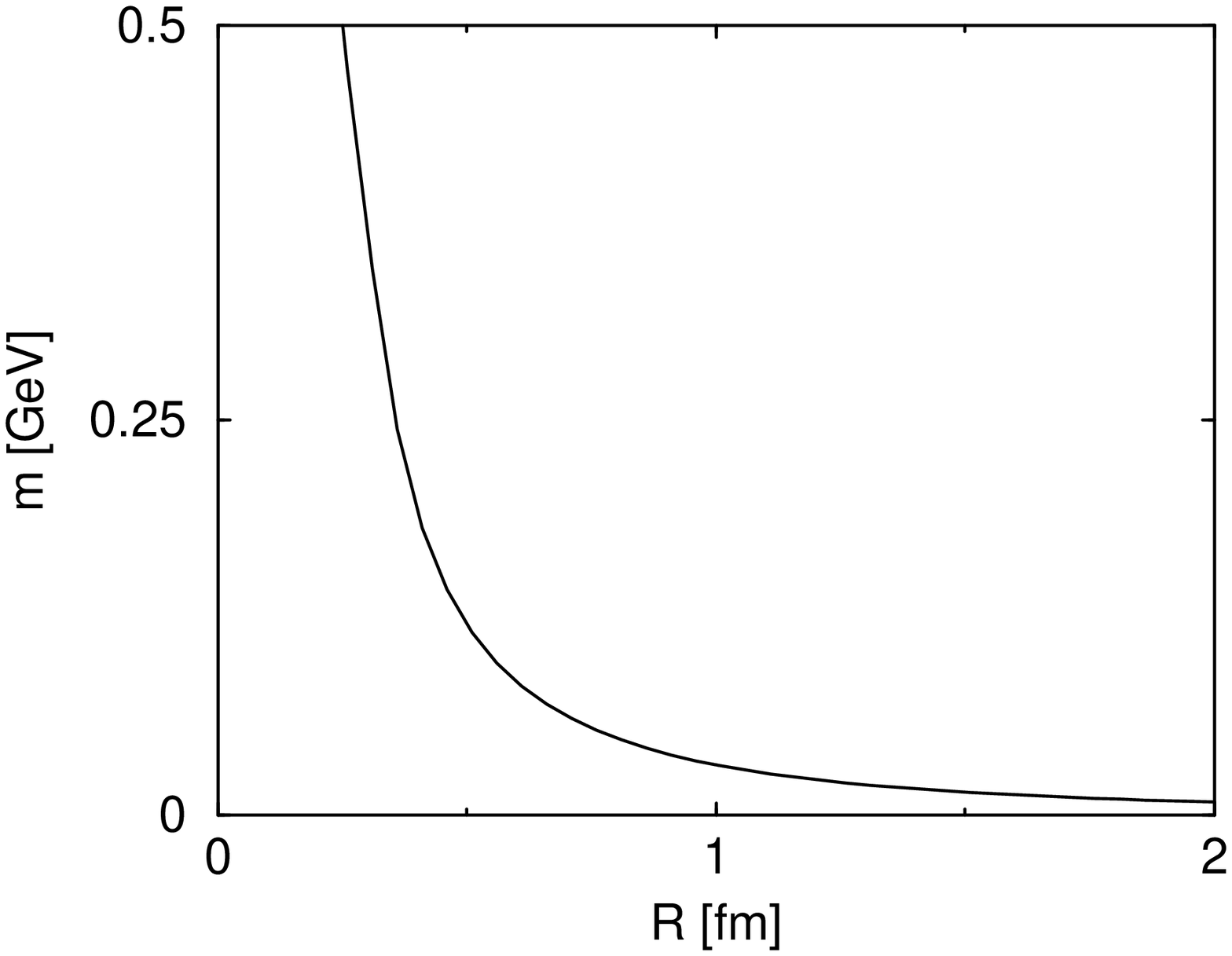}
\end{center}
\vspace*{0.0cm} \caption{\small \baselineskip=0.5cm The quark mass as a function of the bag radius for $G=0.2$.}
 \label{fig : m_R}
\end{figure}

In Fig. \ref{fig : m_R}, we show the dynamical quark mass as a function of the bag radius for $G=0.2$.
The quark mass approaches zero in the large bag radius, which is consistent with the condition of the MIT bag.
On the other hand, the quark mass is dynamically generated for finite bag radii.
Therefore, we find that the NJL interaction induces the dynamical quark mass in the chiral bag.
Although the quark mass becomes too large for small bag radii, $R \ltap 0.2$ fm, it should not be taken too seriously.
For small bag radii, the four point quark interaction $G$ may decrease, hence the quark mass approaches a finite value in the small bag radius limit.
Consequently our model serves a description that confined quarks in a finite size bag behave as constituent quarks.
The fact that the finite dynamical quark mass is induced inside a finite size bag may persist for the realistic situation of 1+3 dimensions, implying spontaneous breaking of chiral symmetry.

%%%%%%%%%%%%%%%%%%%%%%%%%%%%%%%%%%
%%%%%%%%%%%%%%%%%%%%%%%%%%%%%%%%%%

%%%%%%%%%%%%%%%%%%%%%%%%%%%%%%%%%%
%%%%%%%%%%%%%%%%%%%%%%%%%%%%%%%%%%

%%%%%%%%%%%%%%%%%%%%%%%%%%%%%%%%%%
%%%%%%%%%%%%%%%%%%%%%%%%%%%%%%%%%%

\section{Discussion}

One of features in our model is that the quark condensate is induced by the NJL interaction in the chiral bag as a mode sum.
Therefore, it would be interesting to discuss our results in comparison with the previous studies of the non-perturbative feature in the conventional chiral bag model.
In the chiral bag model in 1+3 dimension with massless quarks,
the quark scalar density is shown to take $-0.075 \cos F R^{-3}$ at the center of the bag with the chiral angle $F$ and the bag radius $R$ \cite{Zahed85}.
%\footnote[4]{The chiral angle is indicated by $\theta$ in a notation used in \cite{Zahed85}.}.
This $\cos F$ dependence of the quark scalar density is similar to that obtained in the present calculation, as we see from Eq.~(\ref{eq : massless1}).
Although the direct comparison of the coefficients does not make much sense, qualitative agreement between the two results indicates that we can proceed further discussions about quark condensates in a formalism of the NJL chiral bag model.

Previously, it was discussed by Kunihiro to employ the NJL model inside the chiral bag \cite{Kunihiro1, Kunihiro2}.
There the strong $\Sigma \sim \langle \bar{q}q \rangle$ field is considered to cause the chiral symmetry breaking in the bag.
In our discussion, the dynamical quark mass in the empty bag takes a maximum value at the chiral angle $F=0$ and $\pi$ as shown by the dashed line in Fig.~\ref{fig : mR_F_g} (b).
This means that the quark mass becomes maximum when the point on the chiral circle approaches the sigma axis.
Therefore the idea of Refs. \cite{Kunihiro1, Kunihiro2} is supported by the present analysis in the NJL chiral bag model.

%Lastly we comment on the validity of the mean field approximation in the NJL bag model.
%Generally, it is known that the mean field is not a good approximation in low dimensional systems due to large fluctuations.
%As we see from the Coleman's theorem \cite{Coleman}, the fluctuation in low dimension grows due to the the infrared divergence in the loop contribution and eventually overcomes the mean field.
%However, in the NJL bag model, this is not necessarily the case.
%First, the boundary of the bag cuts off the infrared divergence in the quark loops.
%Second, the finite quark mass has a screening effect for the infrared region in the loop integrals.
%Therefore, the fluctuation will not grow and the mean field approximation may well be valid.

\section{Conclusion}

We discuss the quark condensate in the chiral bag with the NJL interaction inside the chiral bag.
We employ the 1+1 dimensional model in order to avoid numerical complication.
In the outside region of the bag, the sine-Gordon field is introduced for topological properties 
as a pion cloud to mimic the Skyrmion in the 1+3 dimensional system.

Considering the strong correlation between the quarks and pions, the hedgehog ansatz is employed for the mean fields of the quarks and pions.
The scalar and pseudoscalar densities are defined as mean fields with the chiral angle in a self-consistent way.
These quark condensates are then computed explicitly
as mode sums in the chiral bag.
Solutions of the mean field equation leads to the generation of the dynamical quark mass.
As a chiral vacuum polarization, conservation of baryon number is shown to be valid for finite quark condensate.
The chiral Casimir energy is also shown to be well defined without divergence.
There it is important that the symmetry of the single quark spectrum is conserved by the chiral phase.
Consequently, the NJL chiral bag connects smoothly the finite size bag and the Skyrmion as the Cheshire Cat picture implies.
We emphasize that, without the chiral phase, the smooth behavior is not obtained for massive quarks \cite{Yasui3, Yasui4}.

The generation of dynamical quark mass in a bag supports the previous conjectures that the non-perturbative dynamics of quarks and gluons remain inside the bag \cite{Kunihiro1, Kunihiro2, Zahed85}.
It suggests us to reconsider the original picture of the MIT bag model in which a perturbative space is assumed inside the bag.
It would be interesting to consider the relevance to the recent observation in the RHIC experiments which indicate strong quark correlation with finite mass at temperature higher than $T_{\rm c}$.
In order to have clear picture for non-perturbative dynamics of the bag,
it will be an interesting subject to study the realistic 1+3 dimensional NJL chiral bag, and discuss physical observables, which can be  compared with experimental data.

\section*{Acknowledgement}
We express thanks to Prof.~T.~Kunihiro for comments and discussions.
This work is partially supported by a Grant-in-Aid for Scientific Research for Priority Areas, MEXT (Ministry of Education, Culture, Sports, Science and Technology) with No. 17070002.

%%%%%%%%%%%%%%%% Reference %%%%%%%%%%%%%


\begin{thebibliography}{0}
\bibliographystyle{unsrt}

% MIT bag model
\bibitem{MITbag} A.~Chodos, R.~L.~Jaffe, K.~Johnson, C.~B.~Thorn and V.~F.~Weisskopf, {\it Phys. Rev.} {\bf D9}, 3471 (1974).
\bibitem{DeGrand_etal_75} T. DeGrand, R. L. Jaffe, K. Johnson and J. Kiskis, {\it Phys. Rev.} {\bf D12}, 2060 (1975).


% chiral bag model
\bibitem{Chodos} A.~Chodos and C.~B.~Thorn, {\it Phys. Rev.} {\bf D12}, 2733 (1975).
\bibitem{Inoue} T.~Inoue and T.~Maskawa, {\it Prog. Theor. Phys.} {\bf 54}, 1833 (1975).

% little bag
\bibitem{Brown_Rho1979} G.~E.~Brown and M.~Rho, {\it Phys. Lett.} {\bf B82}, 177 (1979).
\bibitem{Brown1979} G.~E.~Brown, M.~Rho and V.~Vento, {\it Phys. Lett.} {\bf B84}, 383 (1979).
\bibitem{Vento80} V. Vento, M. Rho, E. M.Nyman, J. H. Jun, and G. E. Brown, {\it Nucl. Phys.} {\bf A345}, 413 (1980).


% cloudy bag model
\bibitem{Thomas} A.~W.~Thomas, {\it Adv. Nucl. Phys.} {\bf 13}, 1 (1984).

% chiral bag with vector meson
\bibitem{Hosaka90} A. Hosaka, H. Toki and W. Weise, {\it Nucl. Phys.} {\bf A506}, 501 (1990).

% the NJL model
\bibitem{Nambu} Y. Nambu and G. Jona-Lasinio, {\it Phys. Rev}.
{\bf 122}, 345 (1961), {\bf 124}, 246 (1961);
U.~Vogl and W.~Weise, {\it Prog. Part. Nucl. Phys.}
{\bf 27}, 195 (1991);  S.~P.~Klevansky, {\it Rev. Mod. Phys.}
{\bf 64}, 649 (1992); T.~Hatsuda and T.~Kunihiro, {\it Phys. Rept.}
{\bf 247}, 221 (1994).

% quark condensate in the MIT bag
\bibitem{Milton} K.~A.~Milton, {\it Phys. Lett.} {\bf 104B}, 49 (1981).
%
\bibitem{Johnson} K. Johnson, MIT Report No. CTP 1101 (1983).

% A chiral bag model with the NJL interaction
\bibitem{Kunihiro1} T.~Kunihiro, PANIC(1984), Heidelberg, Book of Abstracts I, B41.
\bibitem{Kunihiro2} T.~Kunihiro, Soryuushiron-Kennkyuu, vol. 68, C26 (1983).

% Logarithmic divergence by Debye expansion and quark condensate in a bag
\bibitem{Zahed85} I.~Zahed, A.~Wirzba and U-G.~Meissner, {\it Ann. Phys.} {\bf 165}, 406 (1985).

\bibitem{Kochelev_85} N. I. Kochelev, {\it Sov. J. Nucl. Phys.} {\bf 41}, 291 (1985).
\bibitem{Dorokhov_etal_92} A. E. Dorokhov, Yu. A. Zubov and N. I. Kochelev, {\it Sov. J. Part. Nucl.} {\bf 23}, 522 (1992).

% separation of chiral symmetry breaking and confinement
% chiral soliton model
\bibitem{Birse} M. C. Birse and M. K. Banerjee, {\it Phys. Lett.} {\bf B136}, 284 (1984) .
% chiral quark
\bibitem{Manohar} A. Manohar and H. Georgi, {\it Nucl. Phys.} {\bf B234}, 189 (1984).
% NJL model
\bibitem{Kunihiro85} T. Hatsuda and T. Kunihiro, {\it Prog. Theor. Phys.} {\bf 74}, 765 (1985).

%% AdS/CFT
\bibitem{Bak_Yee} D.~Bak and H.~U.~Yee, {\it Phys. Rev.} {\bf D71}, 046003 (2005).

%
\bibitem{Kiriyama_Hosaka} O.~Kiriyama and A.~Hosaka, {\it Phys. Rev.}
{\bf D67}, 085010 (2003).
%
\bibitem{Yasui1} S.~Yasui, A.~Hosaka and H.~Toki, {\it Phys. Rev.} {\bf D71}, 074009 (2005).
\bibitem{Yasui2} S.~Yasui and A.~Hosaka, {\it Int. J. Mod. Phys.} {\bf E15}, 595 (2006). 
\bibitem{Yasui3} S.~Yasui and A.~Hosaka, {\it Phys. Rev.} {\bf D74}, 054036 (2006).

% instability of little bag without Sjyrme term
\bibitem{Brown_Rho88} G. E. Brown and M. Rho, {\it Commnets Nucl. Part. Phys.} {\bf 18}, 1 (1988).
\bibitem{Hosaka_Toki92} A. Hosaka and H. Toki, {\it Prog. Theor. Phys. Suppl.} {\bf 109}, 137 (1992).
\bibitem{Brown84} G. E. Brown, A. D. Jackson, M. Rho, and V. Vento, {\it Phys. Lett.} {\bf B140}, 285 (1984).
\bibitem{Hosaka_Toki86} A. Hosaka and H. Toki, {\it Phys. Lett.} {\bf B167}, 153 (1986).

% chiral bag model
\bibitem{Mulders} P.~J.~Mulders, {\it Phys. Rev.} {\bf D30}, 1073 (1984).

%% chiral Casimir effect
\bibitem{Hosaka_Toki96} A. Hosaka and H. Toki, {\it Phys. Rep.} {\bf 277}, 65 (1996).

% chiral bag model
\bibitem{Hosaka} A.~Hosaka and H.~Toki, {\it "Quarks, Baryons and Chiral Symmetry"}, World Scientific (2001).

% Dynamical symmetry breaking in asymptotically free field thoeries
\bibitem{Gross_Neveu_74} D. J. Gross and A. Neveu, {\it Phys. Rev.} {\bf D10}, 3235 (1974).

% 1+1 dimensional chiral bag
\bibitem{Zahed84} I. Zahed, {\it Phys. Rev.} {\bf D30}, 2221 (1984).
\bibitem{Zahed_Klabucar} I. Zahed and D. Klabucar, {\it Phys. Rev.} {\bf D30}, 2647 (1984).

% Skyrme model
\bibitem{Skyrme} T.~H.~R.~Skyrme, {\it Proc. Roy. Soc.}  {\bf A260}, 127 (1961).
\bibitem{Adkins} G.~S.~Adkins, C.~P.~Nappi and E.~Witten, {\it Nucl. Phys.} {\bf B228}, 552 (1983).
\bibitem{Zahed_Brown86} I. Zahed and G. E. Brown, {\it Phys. Rept.} {\bf 142}, 1 (1986).

% fractional baryon number
\bibitem{Goldstone} J.~Goldstone and R.~L.~Jaffe, {\it Phys. Rev. Lett.} {\bf 51}, 1518 (1983).

% Logarithmic divergence by Debye expansion
\bibitem{Zahed_Meissner84}  I.~Zahed, U-G.~Meissner and A.~Wirzba, {\it Phys. Lett.} {\bf 145B}, 117 (1984).

% chiral bag with massive quark
\bibitem{Farhi} E.~Farhi, N.~Graham, R.~L.~Jaffe and H.~Weigel, {\it Nucl. Phys.} {\bf B595}, 536 (2001).
\bibitem{Yasui4} S. Yasui, {\it Phys. Rev.} {\bf D74}, 114003 (2006).

% Gell-Mann and Levy
%\bibitem{GeL} M.~Gell-Mann and M.~Levy, {\it Nuovo Cim.} {\bf 16}, 705 (1960).


% Strutinsky method
\bibitem{Wust} E.~W\"ust, L.~Vepstas and A.~D.~Jackson, {\it Phys. Lett.} {\bf B173}, 217 (1986).
\bibitem{Vepstas_Jackson} L.~Vepstas and A.~D.~Jackson,  {\it Phys. Rep.} {\bf 187}, 109 (1990).

% Gaussian
\bibitem{Vepstas_Jackson_Goldhaber} L. Vepstas, A. D. Jackson and A. G. Goldhaber, {\it Phys. Lett.} {\bf 140B}, 280 (1984).

% heat kernel
\bibitem{Hosaka_Toki} A.~Hosaka and H.~Toki,  {\it Phys. Lett.} {\bf 167B}, 153 (1986).

%% 1+1 dim no symmetry breaking
%\bibitem{Coleman} S. Coleman, {\it Commun. Math. Phys.} {\bf 31}, 259 (1973).

\end{thebibliography}
\end{document}